\documentclass[prb,10pt,twocolumn]{revtex4}

\usepackage{graphicx}

\begin{document}

\title{Measurement of Ultrafast Carrier Dynamics in Epitaxial Graphene}

\author{Jahan M. Dawlaty, Shriram Shivaraman, Mvs Chandrashekhar, Farhan Rana, and Michael G. Spencer}
\affiliation{School of Electrical and Computer Engineering, Cornell University, Ithaca, NY, 14853}
\email{jd234@cornell.edu}

\begin{abstract}
Using ultrafast optical pump-probe spectroscopy, we have measured
carrier relaxation times in epitaxial graphene layers grown on SiC
wafers. We find two distinct time scales associated with the
relaxation of nonequilibrium photogenerated carriers. An initial
fast relaxation transient in the 70-120 fs range is followed by a
slower relaxation process in the 0.4-1.7 ps range. The slower
relaxation time is found to be inversely proportional to the degree
of crystalline disorder in the graphene layers as measured by Raman
spectroscopy. We relate the measured fast and slow time constants to
carrier-carrier and carrier-phonon intraband and interband
scattering processes in graphene.
\end{abstract}

\maketitle

Graphene is a single two dimensional (2D) atomic layer of carbon atoms
forming a dense honeycomb crystal lattice~\cite{dressel,nov0}.
It is a zero-bandgap semiconductor with a linear energy dispersion relation for
both electrons and holes~\cite{nov0}. The unusual electronic
and optical properties of graphene have generated interest in
both experimental and theoretical arenas~\cite{nov0,nov1,NovoselovGeimFirst,zhang,heer}.
The high mobility of electrons in graphene has prompted a large number
of investigations into graphene based high speed electronic devices, such as field-effect
transistors, pn-junction diodes and transistors, and terahertz oscillators, and also
into low noise electronic sensors~\cite{NovoselovGeimFirst,lundstrom,marcus,gong,rana,
chemicalsensor}.

The simplest way of obtaining graphene layers is via micromechanical
cleaving (exfoliation) of bulk graphite followed by careful
selection of monolayers by using optical, atomic force, or scanning
electron microscopes~\cite{nov1}. Although this technique results in
relatively high quality films, it might not be suitable for large
scale production. Recently, epitaxial growth of graphene by thermal
decomposition of SiC surface at high temperatures has been
investigated as a promising alternative for large scale
production~\cite{heer,grapheneonSiC1}. This technique can provide
anywhere from a few monolayers of graphene to several ($>50$) layers
on the surface of a SiC wafer. Graphene layers grown by this
technique have demonstrated structural and electronic properties
similar to those of graphene layers obtained by micromechanical
cleaving techniques, including the massless Dirac-like energy
dispersion relation for electrons and holes and carrier mobilities
in the few tens of thousand cm$^{2}$/V-s range~\cite{heer,
grapheneonSiC3,STMgrapheneonSiC}. In addition, the electronic as
well phononic properties of epitaxially grown graphene multilayers
have been found to be different from those of bulk graphite and
similar to those of a graphene monolayer indicating that the
electrons and phonons in different layers in epitaxially grown
graphene are uncoupled~\cite{raman,stacking}. This observed
difference in the properties of epitaxial graphene and bulk graphite
has been attributed to a different stacking scheme for carbon atom
layers in epitaxial graphene compared to bulk
graphite~\cite{stacking}. Epitaxial growth of graphene on SiC
provides a technique to obtain large area multilayers that can be
used for studies, such as ultrafast optical spectroscopy, that are
difficult to conduct on monolayers.

In this paper, we present results from measurements of the ultrafast dynamics
of photoexcited carriers in graphene for the first time. Ultrafast studies of
carrier dynamics in other forms of carbon, such as carbon nanotubes and bulk
graphite, have been reported in the past~\cite{pumpprobeCNT1,
pumpprobeCNT2,pumpprobegraphite1990}. The results presented in this paper are relevant
for understanding carrier intraband and interband scattering mechanisms, and
the corresponding rates, in graphene and their impact on proposed and demonstrated
graphene based electronic and optical devices~\cite{NovoselovGeimFirst,lundstrom,
marcus,gong,rana}.

The graphene samples used in this work were all epitaxially grown on
the carbon face of semi-insulating 6H-SiC wafers using the
techniques that have been reported in detail
previously~\cite{grapheneonSiC1}. Samples A, B and C were grown at
temperatures varying from $1400\,^{\circ}{\rm C}$ to
$1600\,^{\circ}{\rm C}$ and pressures of $2-7\times10^{-6}$ torr.
Micro-Raman spectroscopy of all samples showed a single-resonant G
peak close to 1580 cm${^{-1}}$, a double-resonant D' peak close to
2700 cm${^{-1}}$, and also a relatively low intensity
double-resonant D peak near 1350 cm${^{-1}}$~\cite{raman1,raman}.
The D peak is not allowed in perfect graphene layers since it
requires an elastic scattering process, which is made possible by
disorder, to satisfy momentum conservation~\cite{raman}. The
presence of the D peak therefore indicates the presence of disorder
in the samples. A discussion of the observed intensity of the D peak
and its correlation with the measured carrier relaxation times is
presented later in this paper. Fourier Transform Infrared (FTIR)
spectroscopy of the samples revealed a flat absorption profile in
the entire 2.5-25 $\mu$m range, which is consistent with the
massless Dirac-like energy dispersion of electrons and holes (see
Eqs.~\ref{eq1} and \ref{eq2} below). The number of graphene layers
in samples A, B, and C were estimated from FTIR and XPS spectroscopies
(using the Thickogram method~\cite{thickogram}) to be 6, 12 and 37,
respectively, with less than 5$\%$ error.

\begin{figure}[htb]
\centerline{\includegraphics[width=10cm]{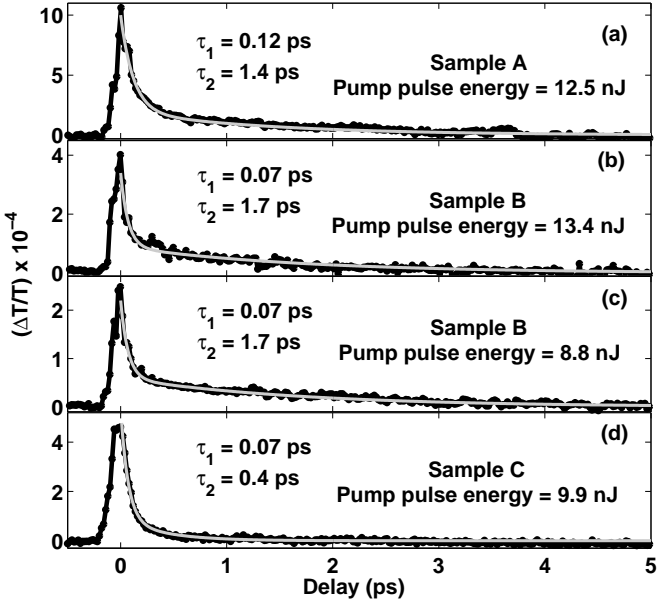}}
\caption{Measured transmittivity transients for (a) sample A (b)
sample B (c) sample B with different pump power, and (d) sample C.
The dark solid lines with markers are the experimental data and the
light solid lines without markers are analytical fits to the data
using exponentials with time constants $\tau_1$ and $\tau_2$. The
transients in (b) and (c) show that the slower time constant
$\tau_{2}$ does not change much as the pump energy is varied.}
\label{sampleAtransient}
\end{figure}

\begin{figure}[htb]
\centerline{\includegraphics[width=8.3cm]{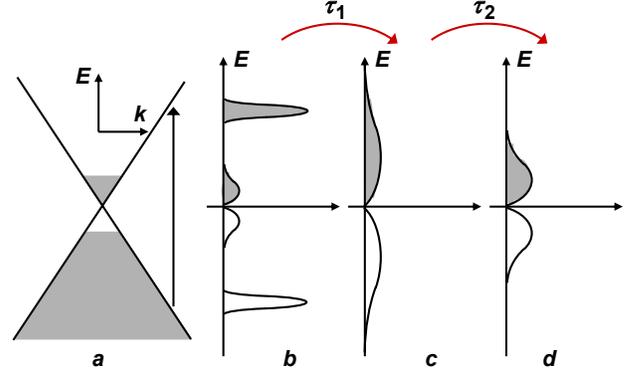}}
\caption{(a)Band structure of graphene showing an intrinsic
population of electrons and holes near the Dirac point. Optical
excitation is indicated by the arrow. (b)The non-equilibrium
distribution of photoexcited carriers account for the initial rise
in transmittivity. (c) The carriers equilibrate among themselves
through carrier-carrier scattering on a time scale given by $\tau_1$
resulting in a hot carrier distribution. (d) Subsequent cooling and
decay of the hot distribution through carrier-phonon scattering (and
possibly electron-hole recombination) occurs on a time scale given
by $\tau_2$.} \label{relaxationdiagram}
\end{figure}

A Ti:sapphire mode-locked laser with 86 MHz pulse repetition rate,
780 nm center wavelength, and $\sim$85 fs pulse width was used for
time-resolved pump-probe spectroscopy of the graphene samples. Pump
pulses with energies between 3-15 nJ were used to generate
photoexcited carriers, while weak probe pulses with energies between
30-100 pJ were used to measure the changes in the transmittivity of
the samples at various delays of the probe pulses with respect to
the pump pulses. The angle of incidence of the pump and probe beams
were $0^{\circ}$ and $15^{\circ}$ respectively. The pump and the
probe were focused to a spot size of about 100 $\mu$m. The
polarization of the probe was rotated by 90 degrees with respect to
the pump and a polarizer was used to eliminate scattered pump light
going in the direction of the probe beam. The probe beam was passed
through a 50 $\mu$m spatial filter for further removal of the
scattered pump light. The pump and probe beams were both modulated
at two different frequencies near $\sim$3 KHz, and changes in the
intensity of the probe pulses at the sum of these two frequencies
were measured with a lockin amplifier.

Photon interband absorption in graphene at optical and near-infrared frequencies
is given by the optical conductivity $\sigma(\omega)$~\cite{rana},
\begin{equation}
\sigma(\omega) = - \frac{e^{2}}{4\hbar}\left[ f_{c}(\hbar \omega/2) -   f_{v}(-\hbar \omega/2) \right] \label{eq1}
\end{equation}
where, $f_{c}(E)$ and $f_{v}(E)$ are the probabilities for the
occupation of an energy level with energy $E$ in the conduction and
valence bands, respectively. The only frequency dependence of
$\sigma(\omega)$ comes from the carrier distribution functions.
Matching the optical boundary conditions at the air/graphene/SiC
interfaces, the optical transmission $t(\omega)$ through $N$
graphene layers on a SiC wafer (normalized to the transmission
through a plain SiC wafer) can be written as,
\begin{equation}
t(\omega) = \frac{1}{1 + N\sigma(\omega)\sqrt{\mu_{o}/\epsilon_{o}}/(1 +n_{SiC})} \label{eq2}
\end{equation}
where, $n_{SiC}$ is the refractive index of SiC. The above expression can be used
to estimate that pump pulses with energies in the range indicated above generate
electron and hole densities in the $3\times10^{11}$-$10^{12}$
cm$^{-2}$ range. The photogenerated carrier densities are larger than the intrinsic
electron and hole densities of $\sim 8\times 10^{10}$ cm$^{-2}$ in graphene at room
temperature.

Figure 1 shows the measured transmittivity transients for different
graphene samples. The figure shows the time dependent change $\Delta
T$ in the transmittivity normalized to the transmittivity in the
absence of the pump pulse. Transmittivity increases sharply just
after photoexcitation. The recovery of the transmittivity exhibits
two distinct time scales; an initial fast relaxation time $\tau_{1}$
in the 70-120 fs range followed by a slower relaxation time
$\tau_{2}$ in the 0.4-1.7 ps range. These time constants have
been extracted by analytical fits to the data using
decaying exponentials. It should be noted here that the
faster time $\tau_{1}$ is of the order of the pulse width and is
therefore not accurately resolvable.

A simple model incorporating band-filling effects together with intraband
carrier-carrier and carrier-phonon scattering can be used to explain the observed
transmittivity transients. Figure 2 is a schematic representation of this model.
As shown in Eq.~\ref{eq1}, optical interband absorption in graphene is proportional
to the difference between the occupancies of the conduction and valence bands at
energies equal to $\hbar\omega/2$ (measured from the Dirac point). Photogeneration
of carriers by the pump pulse reduces this difference and causes the initial
increase in the transmittivity observed in Figure 1. Immediately after photoexcitation,
the non-equilibrium carrier distribution broadens and also equilibrates
with the intrinsic carrier population through carrier-carrier scattering. This
process results in the initial fast relaxation of the transmittivity. The
observed fast relaxation times ($\tau_{1}$) are consistent with the
theoretically predicted carrier-carrier intraband scattering rates in graphene
by S. Das Sarma et. al.~\cite{darma}. As a result of carrier-carrier scattering the photogenerated
carriers are expected to equilibrate among themselves and reach a Fermi-Dirac-like
distribution with a temperature much higher than the lattice temperature.
The observed slower time constant ($\tau_{2}$) of the transmittivity decay  could be
attributed to the subsequent thermalization of the carriers with the lattice
through carrier-phonon intraband scattering.

\begin{figure}[htb]
\centerline{{\includegraphics[width=8cm]{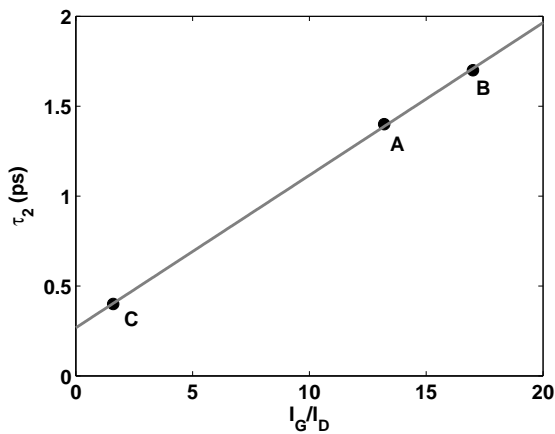}}}
\caption{The slower relaxation time $\tau_2$ is plotted versus the
ratio of the intensity of the Raman $G$ and $D$ peaks for samples A,
B and C. This ratio is a measure of the crystal coherence length.
Larger crystal disorder (smaller coherence length) results in
shorter relaxation times.} \label{tauvsdisorder}
\end{figure}

\begin{figure}[htb]
\centerline{\includegraphics[width=8cm]{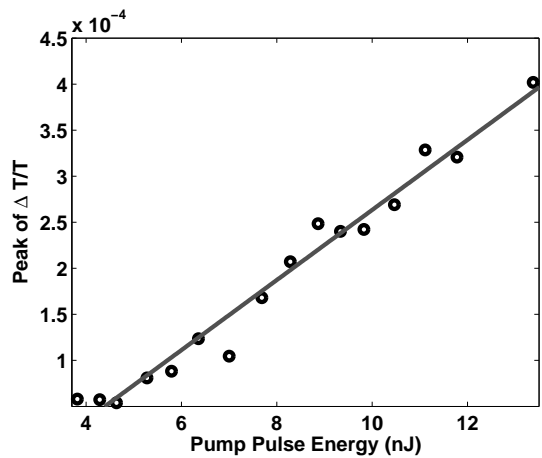}}
\caption{Maximum increase of the transmittivity after
photoexcitation is plotted for various pump energies for sample B.
Linear relation between the maximum transmittivity change and the
the pump pulse energy agrees with the linear absorption and band
filling model and rules out nonlinear two-photon absorption.}
\label{sampleDtransients}
\end{figure}

Electron-hole recombination processes could also contribute to the
slow decay of the transmittivity. The dominant mechanisms for
electron-hole recombination in graphene are not yet well understood.
Electron-hole recombination due to Auger
scattering in graphene was analyzed and carrier density dependent
lifetimes of the order of a few picoseconds for electron-hole
densities in the 10$^{11}$-10$^{12}$ cm$^{-2}$ range were predicted 
by F. Rana~\cite{rana2}. However, in our experiments varying the pump pulse energies in the 3
nJ to 15 nJ range to vary the photogenerated carrier densities did
not lead to any significant changes in the measured values of $\tau_{2}$.
For example, Figures 1(b) and 1(c) show the measured transmittivity transients
for two different pump pulse energies for sample B. These results
indicate that the dominant contribution to $\tau_2$ comes from a process that is
independent of the carrier density and is likely carrier-phonon
rather than carrier-carrier scattering. However, electron-hole
recombination due to carrier-phonon interband scattering cannot be
ruled out. It needs to be pointed out here that carrier-phonon
scattering cannot also be ruled out as a contributor to the fast
relaxation time $\tau_{1}$ from our measurement results. Pump pulse
energy dependence, and therefore carrier density dependence, of the
time $\tau_{1}$ could not be reliably extracted from the
measurements since, as already mentioned above, the observed values
of $\tau_{1}$ were close to the pulse width used in the experiments.

It has been shown that carrier-phonon deformation potential scattering rates
in semiconductor nanostructures are enhanced in the presence of disorder~\cite{disphonon}.
In graphene, the intensity of the double-resonant D peak (near 1350 cm$^{-1}$) in
the raman spectrum can be used as a measure of crystalline disorder since this peak
is absent in perfect graphene layers~\cite{raman,ramandis}. The ratio of the
intensities, $I_{G}$ and $I_{D}$, of the G and D peaks, respectively, in the raman
spectrum has been shown to be proportional to the crystal coherence length~\cite{ramandis}.
Thus, one could expect the measured time constant $\tau_{2}$ to scale with the ratio
$I_{G}/I_{D}$. Figure 3 shows the measured values of $\tau_{2}$ plotted vs the measured
values of the ratio $I_{G}/I_{D}$ for the three samples. Figure 3 shows that
$\tau_{2} \propto I_{G}/I_{D}$ and therefore $\tau_{2}$ is proportional to the
coherence length of the crystal. These results also support electron-phonon
scattering as being the dominant mechanism contributing to $\tau_{2}$.

Figure 4 shows the peak value (normalized) of the measured
transmittivity change $\Delta T/T$ plotted as a function of the pump pulse
energy for sample B. As expected from linear absorption and final
state filling arguments, the maximum value
of  $\Delta T/T$ is proportional to the pump pulse energy. This data also
rules out any significant role played by nonlinear two-photon absorption
in the transmittivity transients. Complete saturation
or bleaching of the absorption is not observed for the range of pump pulse energies
used in the experiments. From the density of states of graphene it follows that
the maximum electron density in an energy interval $\Delta E$ centered at $E$
is $2 E \Delta E/\pi \hbar^{2} v^{2}$, where $v$ is the velocity of carriers in
graphene and equals $10^{8}$ cm/s~\cite{nov0}. Putting $E=\hbar\omega/2$ and
$\Delta E=\hbar \Delta \omega/2$, where $\Delta \omega$ is the optical bandwidth of the
pump pulse ($\sim$10 nm), the maximum electron (or hole) density comes out to be $1.1\times 10^{12}$
cm$^{-2}$ for 780 nm pump center wavelength. The fact that no bleaching of the absorption
is observed even for pulse energies large enough to excite electron (and hole)
densities close to $10^{12}$ cm$^{-2}$ is because of the fast relaxation time $\tau_{1}$
that is of the order of the pump pulse width. Graphene therefore has potential
for use as a fast saturable absorber for generating high energy short pulses from
modelocked lasers~\cite{kartner}.

In conclusion, we have measured ultrafast carrier relaxation rates
in epitaxially grown graphene layers on SiC. We observe two distinct
time scales associated with the relaxation dynamics of
photogenerated carriers. The observed time scales are comparable to
those observed in other related forms of carbons, such as  highly
ordered pyrolytic graphite (HOPG) and single-walled carbon
nanotubes~\cite{pumpprobeCNT1, pumpprobeCNT2,
pumpprobegraphite1990}. Our measurements indicate the separate roles
played by carrier-carrier and carrier-phonon scattering in relaxing
photogenerated carriers.

More work is needed to investigate the role of carrier-phonon
scattering in the fast relaxation time $\tau_{1}$, the role of
electron-hole recombination in the slow relaxation time $\tau_{2}$,
and the value of $\tau_{2}$ in the limit of disorder-free epitaxial
graphene layers. It has been recently pointed out that the
first few carbon layers in epitaxially grown graphene
acquire a bandgap as a result of interaction with the
atoms in the SiC substrate that breaks the symmetry between the $A$
and $B$ atoms in the graphene lattice~\cite{heer2}.
The effect of bandgap on ultrafast intraband and interband
carrier dynamics is not clear. Also, in multilayer graphene
structures the optical response is likely to be dominated
by the large number of layers that are not close
to the substrate and do not have a bandgap.
Pump-probe experiments with fewer monolayers of epitaxial graphene
would be needed to explore the effects of bandgap on carrier
dynamics.

The authors would like to acknowledge support from the National
Science Foundation. This work was also partially supported by the
Air Force office of Scientific research contract No.
FA9550-07-1-0332 (contract  monitor Dr. Donald Silversmith), and
Cornell Material Science and Engineering Center (CCMR) program of
the National Science Foundation (cooperative agreement 0520404).

\end{document}